\documentclass{article}

\usepackage[english]{babel}
\usepackage{eurosym}

\usepackage[usenames]{color}

\begin{document}

\title{Stratified Outsourcing Theory%
\thanks{This work has been performed  in the context of the NWO project
Symbiosis which focuses on software asset outsourcing.
The authors acknowledge Karl de Leeuw, Sanne Nolst Trenit\'{e}, and Arjen Sevenster (all
University of Amsterdam) for discussions concerning outsourcing.}
}
\author{%
  Jan Bergstra,$^{1,3}$
  Guus Delen$^{2}$ \and
  Bas van Vlijmen$^{1}$
\\[2ex]
{\small\begin{tabular}{l}
  ${}^1$ Section Theory of Computer Science,
  Informatics Institute, \\
  Faculty of Science,
  University of Amsterdam, The Netherlands.\\
  ${}^2$ Verdonck, Klooster Associates, Zoetermeer, The Netherlands.\\
   ${}^3$ Department of Computer Science,
 Swansea  University, UK.
  \end{tabular}
}
\date{}
}

\maketitle

\begin{abstract}
\noindent The terminology of sourcing, outsourcing and insourcing is developed in detail on the basis of the preliminary definitions of outsourcing and insourcing and related activities and competences as given in our three previous papers on business mereology, on the concept of a sourcement,  and on outsourcing competence respectively.

Besides providing more a detailed semantic analysis we will introduce, explain, and illustrate a number of additional concepts including:  principal unit of a  sourcement, theme of a sourcement, current sourcement, (un)stable sourcement, and sourcement transformation.

A three level terminology is designed: (i) factual level: operational facts that hold for sourcements including histories thereof, (ii) business level: roles and objectives of various parts of the factual level description, thus explaining each partner's business process and business objectives, (iii) contract level: specification of intended facts and intended business models as found at the business level. Orthogonal to these three conceptual levels, are four temporal aspects: history, now (actuality),  transformation, and transition.

A detailed description of the well-known range of sourcement transformations is given.
\end{abstract}

\section{Introduction}\label{sec:Intro}

In \cite{BV2010} it is argued that in the definitions of outsourcing and related concepts used and surveyed by Delen in   \cite{Delen2005,Delen2007}%
\footnote{%
In these documents a survey is given of definitions of outsourcing, insourcing, outtasking, intasking, follow-up sourcing, back-sourcing, greenfield outsourcing and greenfield insourcing. As early references concerning sourcing Delen mentions: \cite{KernWillcocks1999}, \cite{Lacity1993}, \cite{LohVenkatraman1992}. In addition to these notions multiple outsourcing has become prominent, a definition thereof can be found in \cite{Beulen2000} where multiple outsourcing is subsumed under outsourcing.}
and in many later definitions an ambiguity is somehow built in.%
\footnote{%
In computer science, for some reason, many definitions are deficient. In \cite{Middelburg2010a} it is explained that the experimental process of software testing lacks a definition in the classical literature, and in \cite{Middelburg2010b} it is argued that even for the omnipresent concept of an operating system theoretical informatics provides no informative definition. %
}
An attempt was made in \cite{BV2010} to disambiguate the concepts of insourcing and outsourcing. We will take that paper as the point of departure for this paper. The key design decision of \cite{BV2010} is that outsourcing and insourcing  will refer to transformations rather than to steady states arrived at after an particular transformation.

The reason for disambiguating the terminology discussed in \cite{BV2010} is that very clear meanings are required if one intends to analyze complex sourcing transformations.  In \cite{BV2010} the phrase ``sourcing equilibrium'' has been used to denote a state comprising a number of units, each possibly decomposed into subunits, as well as sources and belongs-to relations, as well as usage relationships between them. However, the phrase sourcing equilibrium is insufficiently flexible to support a
modular approach to
sourcing issues. The term sourcement was coined in \cite{BDV2011a} provides that flexibility. A basic sourcement constitutes of a single source, and a link to its owner, or it constitutes a group of sources which will not evolve into divergent ownership. In this paper we will work out a number of consequences of the decision to disambiguate outsourcing in this particular way. We refer to \cite{BDV2011a} for a general
discussion of the concept of outsourcing at large. What can and what can't be outsourced in most general terms is studied in that work in some depth.

In \cite{BDV2011b} we work under the assumption that sourcing and outsourcing stand for activities
which take place on a large scale in today's economy. Closely  related to sourcing and outsourcing is the repertoire of activities of the service industry. In \cite{BDV2011b} a proposal is made on how to look at
the concept of competence for these different areas. The paper proposes that, for instance, outsourcing competence must be viewed as a community competence. Further the point is made that outsourcing
competence must not be subsumed under sourcing competence, because sourcing is not a generalization of outsourcing, its objectives are different to a significant extent.

Outsourcing theory must be considered as a tool for strengthening that community
competence. The work in \cite{BDV2011b} sets the stage for this theory by investigating the requirements for it. Development of outsourcing theory should preferably be
guided by community competence pull, rather than that an autonomous theory
push must be organized for propagating results of theoretical research into
practice. It is our ambition that the outsourcing theory that is worked out in
the sequel can be perceived in that light. A second design principle
has been to insist on finding non-circular definitions for key notions concerning outsourcing. We hold that circular definitions are very unhelpful if the terminology is used to understand and explain what goes on
in some business process, rather than merely to provide distant and abstract
specifications.

One reason for carrying out this work has been the first author's experience
in the University of Amsterdam's works council the other motive being our
commitment to carrying through the Symbiosis project on software asset outsourcing.

In the mentioned works council time and again outsourcing processes
need to be discussed and for unclear reasons the terminology of outsourcing
never quite works, a consequence of that failure being that significant details remain
unrecognized or unanalyzed only to emerge as decisive in a too late stage.
That raises the question how to develop a way of communicating about these
matters which helps to avoid surprises?

The design of a theory of outsourcing
is an elaborate exercise which by no means comes to an end with this paper.
On the contrary, many directions for useful further work can be spotted.

\subsection{Outline of the paper}
We will briefly explain the plan of  this paper. In Section \ref{ACCA} the connection of this work with \cite{BDV2011b} is made explicit. In Section \ref{Stratif} stratification of outsourcing related concepts is introduced, and in  \ref{sourceToUnit} a range of technical notions is introduced that allows for precise descriptions of the relative status of sources and units. In Section \ref{ThreeLayers} three different layers of description of sourcements are introduced and explained. This
sets the stage for a stratification. In Section \ref{FTA} four
aspects of timing are put forward and a thereafter in Section \ref{attributes-of-configurations} a
survey of attributes of all entities involved is provided. In Section \ref{Oid} outsourcing is discussed in thorough detail,
and in Section
\ref{Ofstid} the other fundamental sourcement transformations are discussed. Subsequently in Section \ref{Trc}
the notion of a lot is analyzed as a method for specifying functional requirements for a sourcement transformation. We end wist some conclusions in \ref{Conc}.

\section{Audience, competence, and conjectural ability}\label{ACCA}
This work below will be based on \cite{BDV2011b} which contains a listing of requirements which is supposed to be met by a forthcoming outsourcing theory. Familiarity with \cite{BDV2011b} will be assumed. The audience of the proposed theory consists of those individuals who have a
so-called outsourcing framework competence as specified in \cite{BDV2011b}.
The theory is advocated by the group of its authors who as it stands have not achieved anything like a community confirmation of the status of the work. It follows that abilities
acquired by persons taking notice of the outsourcing theory of this paper are by definition non-community confirmed conjectural abilities. The requirements of \cite{BDV2011b} state that it must be indicated what conjectural abilities are
supposed to be acquired through becoming familiar with the theory. We will deal with that matter in this Section only, by listing a number of conjectural abilities that we hope contribute to by means of the theory to be outlined in the rest of the paper.

\subsection{Ability to produce detailed current state descriptions}
The stratified model is supposed to strengthen one's ability to provide precise definitions of the current state
of a sourcement configuration involving one or more units. The detailed description of fundamental sourcement
transformations calls for taking into account many aspects that might otherwise be overlooked. Using the degree of outsourcingness as proposed in Section \ref{DOO} it can be estimated to what extent some aspects need to be taken into account.

This conjectural ability to provide precise and  complete descriptions can be helpful in the following circumstances:
(i) when deciding whether or not the application of EU procurement can be avoided, (ii) when deciding to what extent
service procurement (and  service science) provides a more productive perspective than  outsourcing theory, (iii)
when designing the level of abstraction at which an intended sourcement transformation is best specified in advance of procurement (often called vendor selection),
(iv) when designing a transition that implements a given sourcement transformation.

\subsection{Ability to search for similar cases}
The terminology of the proposed outsourcing theory allows to classify intended transformations with a high
resolution. That allows one to search for similar cases effectively.
Instead of almost randomly collecting data concerning
past instances of sourcement transformations one may try to find similar  cases which have been brought to completion in the past. Such transformations can be studied in detail in order to assess the effectiveness of the
course of action then taken, with the intention to determine best practices for the case at hand
and to avoid mistakes which are know to have occurred in previous similar cases.

\subsection{Ability to design complete transition plans}
Making use of detailed descriptions of initial, intermediate, and final, sourcement configurations it is possible (conjectured) to design life-cycles for sourcement tranformations from which multi-threaded algorithms for effecting transitions implementing  the intended transformations can be inferred.

\subsection{Ability to turn a competence into a focused capability}
By making use of detailed and comprehensive methods of classification a person can turn a flat listing
of experiences with roles in various stages of sourcement transformations into a more informative picture. Whereas this  listing may suffice to demonstrate significant community confirmed outsourcing competence a further
classification may establish the person's acquaintance with a range of different circumstances all related to either some specific domain or to some kind of transformations, perhaps but not necessarily restricted to a specific domain.
This ability may be helpful for playing a role in task acquisition as well as for adequate role assignment within a
task.

\subsection{Ability to collect meaningful data}
The ability to provide a high resolution classification of sourcement transformations also generates a
(conjectural) ability to extract data from current transformation cases and to store these in a structured fashion. This ability will not immediately apply when a new sourcement transformation and its implementing transition plan need
to be designed or a new sourcement procurement is sought, but it represents a precondition for collecting data in such a way that on the long run evidence based best practices can come into existence.

We assume that the conjectural abilities listed in the previous items are all evidence respecting (ERCA in the sense of \cite{BDV2011b}), and some are evidence oriented (EOCA in the sense of \cite{BDV2011b}). For that reason some efforts of persons involved in decision making on sourcement
transformations towards data collection that eventually prepares an influential evidence base may be
expected, so that an ability in that direction is welcome, even if it does not qualify as an outsourcing ability rightaway.

\section{Stratification, sources, and units}\label{Stratif}
Given the definitions of insourcing and outsourcing provided in \cite{BV2010}, and the notions of a sourcement, basic sourcement and sourcement portfolio, as introduced in \cite{BDV2011a}, still many issues and questions concerning sourcing theory remain, some which have been listed in \cite{BDV2011b}
as requirements for an outsourcing theory. We will not repeat those issues here. In general, however, these questions indicate that the jargon may yet be lacking the necessary unambiguous semantics.  A problem which' solution requires both the development of new terminology and working out better assignments of meaning to known terms and phrases.

\subsection{Stratified concepts: housing as an example}
By making use of a stratification of concepts circular
dependencies between them will be prevented.  Following \cite{BDV2011b} we will use housing as a running example
for explaining the purpose of stratification.

If $A$ inhabits home $H$ owned by $B$ (factual level information), then more specifically $A$ may be renting $H$ from $B$ (business level information). That fact may be regulated by means of some contract $C$ (contract level information). $A$ may expect to leave $H$ (transformational aspect information) at time $t$ because this regulation occurs in $C$.

We notice that the assertion that $A$ is renting $H$ makes sense because $A$
inhabits $H$, (not because of the existence of a contract which then would lead to a circularity), whereas how to obtain modifications of the renting process is regulated by  contract $C$. Neither the contract, nor the phenomenon of renting are needed, however, to grasp the facts regarding $A$ living in $H$ and of $H$ being in the possession of $B$.

This example clearly indicates the three level, or three dimensional, stratification which we propose to use for sourcing and outsourcing, and more specifically for the description of sourcements: (i) facts on the ground, (ii) business processes, and (iii) contracts.

Of course the existence of a contract is itself a fact, but the content of the contract need not comply with the observed facts of the moment. Thus in level (iii) it is the meaning, both intended and unintended, of the contracts which is meant. If a contract has been signed, but the document has been lost, that impacts on the facts on the ground (paperwork missing), but it won't necessarily change the contract configuration. The
circumstance that some facts are caused by the existence of contracts and by the business opportunities created by way of those contracts does not contradict the independence of the facts from the contracts. At
any moment of time one may ask to what extent the facts on the ground and the business processes making use of those facts comply with the configuration of contracts, both formally and intentionally.

Circularities  may render large fragments of theory essentially meaningless and that strikes us as being
most important.  The proposed stratification is primarily helpful for getting potential circularities out of the way.
Removing circularities often requires the introduction of new or unusual aspects which support a
hierarchical configuration that was absent or invisible before. Thus the focus of this paper will be on non-circular definitions and specifications. The key ingredient for that is to develop a hierarchically layered, that is stratified, structure for the relevant concepts together with a terminology that allows one to be precise
about aspects relevant to a certain layer without even mentioning any aspects of higher layers.

In the renting example three levels of description can be disentangled and non-circular definitions of the underlying concepts can be given: inhabiting $H$ is not dependent on any contract configuration and so on.

\subsection{A plurality of units}
We imagine that the world or a relevant part of the world which is under consideration consists of a plurality of units containing a plurality of sources, each of which is wrapped in a basic sourcement. The family of
basic sourcements used for realizing a theme maintained by $U$ is called a sourcement. A substantial and coherent subfamily of a sourcement may also be referred so as a sourcement (rather than merely some
family of basic sourcements). The combination of a collection of sourcements used by a unit $U$ for various
of its themes may be called a sourcement portfolio. A sourcement portfolio is just a union of sourcements equipped with a one-level hierarchical structure.

At some instant of time a snapshot
of the survey of units and sources and their usage is called a global sourcing state. This world is constantly being transformed by sourcing transformations. If one thinks in terms of basic sourcements,
sourcements and sourcement portfolios the larger aggregation level one considers the shorter the duration of stable periods. In an organization of several thousands of employees it is likely nowadays that
its overall sourcement portfolio is always (somewhere) in the process of being changed. An equilibrium
at the level of temporary invariance of sourcement portfolios may never exist.

However, there may be an equilibrium at a higher level of abstraction, like in theoretical micro-economics where a Walras equilibrium indicates the
presence of equilibrium prices and not the presence of stable ownership and or usage relations between
agents and goods.

Thus speaking of a global sourcing equilibrium, if it is workable at all, requires a perspective at a high
level of abstraction in such a way that some dynamics is still possible given an equilibrium state. A sourcing architecture provides information about sources and units which is considered normative for an equilibrium.

Now if one restricts attention to a very limited number of units and their respective sources,
the projection of a global sourcing state to that limited part of the unit class will show
relatively few transformations per unit of time. Thus when restricting attention to a part of the collection of units a sourcing equilibrium may be observed at the lowest level of abstraction. In that equilibrium the use relations between units and sources are stable during a significant time interval.

\subsection{Source to unit relations}\label{sourceToUnit}
To begin with, a more detailed terminology is outlined for describing notions  needed  in addition to what was introduced in \cite{BV2010}.
These notions help to describe the variety of possible relations between sources and units.
\begin{description}\label{terminology}
\item{\em Theme.} Units maintain (keep, manage, elaborate) themes.\footnote{Themes don't occur in \cite{BV2010}. Different units may maintain themes with identical names. Such themes are formally distinguished by always tagging (perhaps in one's mind only a theme with the unit by which it is being maintained).} A theme represents an objective or package of objectives of a unit. Performance for themes is delivered by using single or cooperating sources. The functional relation between source and unit must always be understood via a theme.\footnote{In/outsourcing transformations leave themes and their maintenance invariant. Themes may be considered components of a unit's mission.  Themes are needed to express mission invariance.}

\item{\em Theme cluster.} A cluster of themes is a set of themes maintained by the same unit for which it makes sense that these are maintained simultaneously in a coordinated fashion.\footnote{Selling bikes, buying used bikes, and bike repair may be a cluster of  themes, while selling specialized wear for biking is another theme constituting a group of its own.}
\item{\em Owner.} Sources always have owners, which are either units or subunits.
\item{\em Use.} Unit $U$ uses a source $S$ if $S$ is used for some theme $T$ which is maintained by $U$.\footnote{If $U$ uses $S$ for theme $T$ it will alternatively be said that $S$ supports $U$ for $T$. If $U$ is known from the context we also say $S$ supports $T$.}
\item{\em Selfsourcing for source.} Unit $U$ is selfsourcing for source $S$ if $U$ uses $S$ and $U$ owns $S$ at the same time.\footnote{Selfsourcing may also be termed intra-unit sourcing.}
\item{\em Non-selfsourcing for source.} Unit $U$ is non-selfsourcing for source $S$ if $U$ uses $S$ and $S$ is not
owned by $U$ at the same time.\footnote{It is common to allow `$U$ is non-selfsourcing for source $S$' as a possible meaning of `$U$ is outsourcing $S$' but we will not to do so. That
follows from letting outsourcing denote a transformation rather than a state \cite{BV2010}. Non-selfsourcing for $S$ may also be termed inter-unit sourcing of (for) $S.$}
\item{\em Source type.} Every source is member of a source type. Different units may own sources of the same type. A unit may own any
number of sources of a given type.
\item{\em Singleton source type.}  Singleton source types allow at most one source of that type to be owned by a unit.
\item{\em Selfsourcing for type.} Unit $U$ is selfsourcing for source type $\tau$ if every source of type $\tau$ which is used
by $U$ is  owned by $U$ and if moreover $U$ uses at least one source of type $\tau.$
\item{\em Partial selfsourcing for type.} Unit $U$ is partially selfsourcing for source type $\tau$ if one or more sources of type $\tau$
which are used by $U$ are also owned by $U.$%
\footnote{%
By definition if $U$ is selfsourcing for type $\tau$ then $U$ is partially selfsourcing
for type $\tau$. The notion of partial selfsourcing makes sense for source types only.%
}
\item{\em Non-selfsourcing for type.} Unit $U$ is non-selfsourcing for source type $\tau$ if $U$ uses at least one source of type $\tau$ and for every source of type $\tau$ which is used by $U$ is not owned by $U$.
\item{\em Partial non-selfsourcing for type.} Unit $U$ is partially non-selfsourcing for source type $\tau$ if at least one source of type $\tau$ which is used
by $U$ is not owned by $U.$\footnote{By definition if $U$ is non-selfsourcing for type $\tau$ then $U$ is partially non-selfsourcing for type $\tau.$}
\item{\em Dependent sources.} Sources can be mutually dependent. For instance a machine needs to be placed in some premises, positioned at some location, maintained by
a certain crew and operated by another team. That situation gives rise to five sources which need each other's existence in order to be used
productively.%
\footnote{Subsets of these sources can be understood as a single source, however, as long as transformations deal with their members in a similar and compatible way. Decomposition of sourcements may be needed in advance of a transformation. Whether or not an outsourcing transformation will involve a current sourcement decomposition may not be known in advance. Indeed it may not be known before the decision to engage in outsourcing the sources involved in that current sourcement has been made.}
\end{description}

\subsection{Consequences by way of examples}\label{examples-additionaltermi}
Some simple consequences can be derived from the definitions just given.
These consequences can serve as examples of the use of the notions as well.

\begin{enumerate}
\item After a source $S$ supporting theme $T$ originally belonging to unit $U$ has been outsourced by $U$ to some different unit $V$ then $U$ is {\em non-selfsourcing for} $S$.\footnote{The non-selfsourcing status will not continue indefinitely.  From \cite{BDV2011a} we repeat the following comparison. One may compare outsourcing with diving: after the dive there is the temporary status of being submerged. Diving has an unresolved ambiguity comparable to that of outsourcing.}
\item Suppose unit $U$ maintains theme $T$ and is self-sourcing for source $S$ which supports $T$. We consider the situation that after a single sourcement transformation $U$ is non-selfsourcing for $S$. It is now adequate to conclude that $S$ has been outsourced to some other unit.
\item However, if from the same initial situation a state is reached where $U$ is non-selfsourcing for $\tau$ with $\tau$ the type of $S$ and
$U$ owning only a single source of type $\tau$ then it cannot be inferred that $U$ has outsourced $S$. It may have dropped $S$ altogether and may now be using a different source $S^{\prime}$ of type $\tau$ owned by another unit, say $V$.
\item In addition to the terminology proposed in  \cite{BV2010} about unit $U$ outsourcing source $S$, one may distinguish $U$ outsourcing source type $\tau$ which consists of a transformation after which $U$ is non-selfsourcing for $\tau$.\footnote{An outsourcing of type $\tau$ by unit $U$ need not involve an outsourcing of all sources of type $\tau$ by $U$. Once `outsourcing of type' has been
introduced `partial outsourcing of type' also emerges as a plausible notion, which leads to a state of partial non-selfsourcing. Application of partial outsourcing can be avoided by splitting the type in subtypes before outsourcing.}
\item After $U$ has outsourced source $S$ to unit $V$ it can be backsourced again. That backsourcing is also an instance of insourcing. Backsourcing is an insourcing of source, not an insourcing of type. Backsourcing is `undefined' after an outsourcing of type.
\item Outsourcing followed by backsourcing leads to (a state of) selfsourcing.
\end{enumerate}

\section{Dimensions of status}\label{ThreeLayers}
The language for talking about sourcements which we are in part developing here  should be effective during discussions about sourcing and sourcing transformations. Often contracts are not
yet known, transformational processes have still to be determined and an adequate description of the current sourcement is
needed in the absence of a firm view on expected future sourcements. It must be possible (facilitated by the terminology and its method
of use) to state and answer the following generic questions in a particular case:
\begin{itemize}
\item Is the sourcement consistent with a given business view for the units involved? (This cannot be analyzed if the sourcement is defined in terms of the business view.)
\item Is the sourcement plus business view consistent with a contractual description? (This defeats analysis if the contracts enter the business description.)
\item Will a sourcing transformation lead to a sourcement reflecting a specific (intended) architecture?
(This cannot be properly analyzed if the target sourcement portfolio can't be specified otherwise than
in terms of the architecture.)
\end{itemize}

\subsection{Three dimensions of status}
The stratification proposed begins with making a distinction between three levels, or dimensions. These different levels will return whenever a state of affairs, such as a sourcement or a sourcement portfolio
must be discussed, specified, designed, critically assessed, or simply criticized.
\begin{description}
\item{\em Fact.} The facts on the ground describe what activities take place by whom and which sources are used.
\item{\em Business.} The business level provides objectives and roles of participants of the activities supporting a theme in a quantified fashion.
\item{\em Contract.} The contract position of various actors provides  a regulated setting from which all parties involved may infer their rights and expectations.
\end{description}

A status at each level is captured by an appropriate configuration. By default a sourcement will be identified
with its configuration of facts on the ground. But sourcements have both other dimensions as well and
when necessary that must be taken into account.

\begin{description}
\item{\em Sourcement.} The definition has been given in \cite{BV2010}.
\item{\em Business configuration.} Often called the business model. This model must be provided for each unit participating in a sourcement.
\item{\em Contract configuration.} Often called contract position. Specification of the business configuration by means of contracts (promises, covenants, agreements, treaties, letters of intention, letters of understanding).
\item{\em Transformation configuration.} A bundle of transformations for one or more sources mentioned in a sourcement.
\end{description}

\subsection{Informal explanation of the three layers}
In order to further clarify what aspects are covered in the different levels of the three level model we will phrase the matter in different words. The hierarchy of layers, that is proposed with the objective of organizing concepts in such a way that the risk of circularities vanishes, is informally worded as follows:
\begin{description}
\item{\em Facts on the ground.} What one can see, what is measured and quantified.
\item{\em Business rationale: business models and business cases.} Why things are done in a
steady state (that is a sourcement equilibrium) in the way they are.
\item{\em Contracts, promises, agreements and legal position.} Why things are changed (or left unchanged) the way they are, insofar as business rational is not driving the changes or stabilities.
\end{description}

\noindent `Facts on the ground' is the lowest level of description. Non-circularity is achieved by having the facts on the ground described independently from business rationales, contracts and sourcing transformations, and by having the business rationales specified independently from contracts and transformations.

Undeniably some business rationales may have been
formed because of a potential support from known contract
types. We will hold, however, that in the first order business rationales can and must explain
contracts rather than the other way around.

\subsubsection{Ways of speaking about operational facts on the ground}
All forms of specification are valid in this case and there is no need to either review or enrich the language needed for that purpose.
\begin{itemize}
\item Facts on the ground include first and for all the primary streams of money, the transactions and operations, the various management and reporting configurations. Of course descriptions of these are always partial and incomplete. In many cases it will suffice to provide descriptions that allow to determine which of a unit's activities and sources are considered part of a particular sourcement.
\item Operational facts include monitoring and assessment activities: customer satisfaction analysis, quality assessment and assurance,
assessment of the  portfolios of licenses and certifications.
\item When describing these facts it may be important to be explicit about concurrent activity within or between various units. We mention the thread algebra of \cite{BergstraMiddelburg2007} as a very simple model of concurrent activity which may be used for these purposes. In \cite{BaetenBastenReniers2009} one finds process algebra which proves a far richer language for the specification of concurrent activity.
\end{itemize}

\subsubsection{Ways of speaking about business rationale and business case}
It is important that each description in the business level category contributes to the understanding of the way in which sources and operations of a unit that participates in a sourcement constitute a part of its business. If something is owned: how are the costs of ownership measured and covered? If something is rented or leased: what is the model of that transaction? And so on.

Reference to contracts must be avoided by all means because these descriptions may be needed for instance to detect mismatches with what is said in the contracts, or to help with designing an adequate package of contracts in case the operations are already up and running. Here are some aspects that may be covered in the business layer.

\begin{itemize}
\item The business configuration covers an explanation of how the operations of various units contribute to their business. For instance if premises are rented the rent is cost, while the revenues of operations contribute to profits. It must be explained how, at all, a profit can be made, both by the principal unit and by the providing units. It must be explained what should be maximized, minimized, or optimized in order to run an adequate business and it should be explained which transformations are envisaged to improve that business when needed.
\item In the business layer it is explained who are the customers of each service and how the needs and wishes of the customers are measured and taken care of. It must be explained who carries responsibility for various forms of quality of service involved and how these responsibilities are taken care of.
\item Business information will indicate for a source that it is exploited with some objectives in mind as well as with a number of constraints.
\item At this level one expects an explanation of the way personnel has been
hired (permanent staff, temporary staff, employed staff, not employed staff, through an intermediate agency), in terms of expected advantages now as well as degrees of freedom (or also advantages) in the future.
\item For all money streams it must be clarified whether or not the stream contributes to an understanding of the
business,\footnote{If parking is made expensive merely to regulate parking capacity the corresponding money stream may not even contribute positively to the business of parking place management.} or how it constitutes part of an essential feedback control loop.
\end{itemize}

\subsubsection{Ways of speaking about sourcing related contracts}
Describing the contract position may but need not involve the copies of contracts per se. Important is that is indicated what aspects are contractually laid down and if that can be expressed in an informative way more abstractly than by means of presenting the actual contracts in full text that may be used as a description.
Many matters are covered by contracts only. For instance a contract may require that a unit maintains an adequate portfolio of licenses and certificates for its sources for each of the services it is delivering (for its own unit or for other units).

\subsection{Three levels of description}
When writing or speaking about sourcing phenomena one needs sourcement descriptions rather than sourcements themselves. The three layers of status translate into our layers of description.

\begin{description}
\item{\em Sourcement descriptions.} A sourcement description specifies the facts on the ground. These include historic information about preceding sourcements and corresponding business descriptions and contract position descriptions.
\item{\em Business configuration description.} The business configuration description corresponding to a sourcement provides information about how the business is viewed. For instance: what is maximized or optimized, what counts as cost, who is in control which objectives govern the operations of the sourcement, who is the (real) customer.
\item{\em Contract configuration description.} The contract configuration description explains the various legal positions as well as their histories.
\end{description}

These species of description are mutually disjoint. Although it may be tempting to consider a contract a fact on the ground doing so puts the comprehensibility of  all descriptions and documents at risk.

\section{Four temporal aspects}\label{FTA}
Aspects involving time at the time scale of changing sourcements is called a temporal aspect. Four temporal aspects will be distinguished:
\begin{description}
\item{\em History.} Historic information comes in two flavors. The first being a specification of a state of affairs in past time. In the context of outsourcing the ownership history of a source is important, and the development history of a contract, and of course the business history of various units. Historic information can be assembled and stored without a focus on what was known by whom at various moments of time. The second flavor takes
one or more units (or agents within units) and some moment in time as a point of sight and it provides a picture of the history of various relevant units, themes, sources, service deliveries, service consumptions, business cases, business actions, agreements and contracts as know to the mentioned agents.
\item{\em Now.} Descriptions of specifcations expressed in terms of ``now'' are a tool for specifying a context
from which decisions about the future course of actions can be taken. Historic information must be cast in terms of what is known now, and the future is at best known in terms of intentions, promises, expectations, or probabilities. If a decision is assessed in hindsight one must design an adequate picture of the actual situation as it was seen by participants.

Thus `now'  can appear in two different roles: as a flag indicating that a description is meant to serve decision making at this very moment in time, and as an indication that some past time is temporally referred to as now in order acquire an adequate picture of the scene of action then (of course implicitly
assuming that then, present time was referred to as ``now''.)
\item{\em Transformation.} Transformation is about the specification of goal sourcements which are or might be preferred.
\item{\em Transition.} Transition deals with the path of actions and intermediate states which implements
an intended transformation.
\end{description}

\subsection{Description of temporal aspects}
Except for ``now'' the other temporal aspects are accessible via descriptions only. Transformations need
a description of now and of a state in future time. Transitions, however, need a specification of a path or course of action. For sourcements it is most easily imagined what the differentiation of temporal modalities
leads to in terms of useful further qualifications of a sourcement:
\begin{description}
\item{\em Current sourcement.} The current sourcement is transformed into another current sourcement as the result of outsourcing and insourcing.
\item{\em Past sourcement.} The past sourcement precedes a particular sequence
sourcing transformation.
\item{\em Sourcement history.} This history refers to the sucession of sourcements containing a particular source. Information about the sourcement history may be needed
to fully understand the legal positions of employees constituting (or belonging to) a source that has gone through different sourcing transformations.
\item {\em Future sourcement.} A future sourcement specifies the intended current sourcement after a forthcoming transformation has taken place.
\item{\em Presourcement and postsourcement.} Following the programming terminology of precondition and postcondition,
we may label the current sourcement which is to be transformed as a presourcement and the intended future sourcement that is to result from a transition as a postsourcement. During a transition there may be intermediate sourcements.
\end{description}

\subsubsection{Ways of speaking about outsourcing histories}
Facts on the ground can be examined without much consideration of business cases or of contracts.  A sourcing architecture history is considered a constituent of the facts on the ground.

Suppose unit $U$ maintains theme $T$ and source $S$ is used in order to carry out thematic operations $O_T$ for theme $T$. Thus $U$ is the principal unit with theme $T$ and makes use of source $S$.

We will discuss this state in more detail and in the following description $U, U^{\prime}$,
and $U^{\prime\prime}$ represent pairwise different units.

Assuming the above state of affairs the following possibilities can be distinguished:
	\begin{itemize}
	\item Source $S$ is internal to $U$ (alternatively phrased: $U$ is selfsourcing w.r.t. $S$ for theme $T$) and
	\begin{itemize}
	\item either $S$ has been developed inside $U$ and has not been outsourced by $U$ since $S$ came into existence.
	\item or $S$ has been insourced from some $U^{\prime}$ in the most recent sourcing transformation involving $S$, or
	\item $S$ has been follow-up outsourced by some third unit $U^{\prime\prime}$ which previously had outsourced it to $U^{\prime}$ in the most recent sourcing transformation involving $S$, or
	\item $S$ has been backsourced to $U$ after having previously been outsourced by $U$,
	to some other unit $V$, or in a chained fashion to a sequence of units.
	\end{itemize}
	\item Source $S$ is external to $U$ (alternatively phrased: $U$ is non-selfsourcing w.r.t. $S$ for theme $T$), in which case
	source $S$ is internal to some different unit $U^{\prime}$ (alternatively: $U^{\prime}$ is selfsourcing w.r.t.\ $S$ for
	providing operations $O_T$ concerning theme $T$ of unit $U$). In this second case:
	\begin{itemize}
	\item either $S$ has been outsourced from $U$ to $U^{\prime}$ in the most recent sourcing transformation involving $S$, or
	\item $S$ has been follow-up outsourced by $U$ from a third unit  $U^{\prime\prime}$ to $U^{\prime}$ in the most recent sourcing
	transformation involving $S$, or,
	\item $S$ has been developed inside $U^{\prime}$ and has not been outsourced by $U^{\prime}$ since it came into existence.
	\item $S$ has been developed inside $U^{\prime\prime}$ and has been successively  outsourced and follow-up sourced by a succession
	of sourcing transformations ending within $U^{\prime}$.
	\end{itemize}
	\end{itemize}

\subsubsection{Ways of speaking about sourcing transformations}
Sourcing transformations can be specified on three levels: facts, business and contracts.
At each level several different methods to describe a past, ongoing, or planned, transformation
can be considered, all derived from similar issues in system design and software technology.
We provide some brief comments only about this matter.

Pre and post condition pairs can be used to specify the state before a transition and the intended state after a transition, while leaving open how the transformation itself may work.

An intended transformation is obtained by putting into effect a transition plan. Given an abstract description of the transformation, for instance in terms of presourcements and postsourcements, or simply in terms of preconditions and postconditions, an implementing transition plan must be designed.%
\footnote{Temporal logic may be used to add to this description more information about intermediate states visited during a transformation. An abstract  program or process notation can be
used to specify the order of steps that suffices to effect the transformation, in which case
pre and postconditions serve as a more abstract (extensional) specification and verification of the intensional (algorithmic) specification against the extensional specification
becomes possible in principle.}

The business case for an outsourcing transformation is the listing of
business cases for each participating unit as given by comparing the business expectation in the expected sourcement of that unit with its current sourcing aggregation, while ignoring the cost of transition. The inclusive business case of a sourcing transformation complements its business case with expected costs and risk analyses for the intended transition
plans of each involved unit.

Business cases can be specified without much consideration of these facts on the ground. At closer inspection we must distinguish a {\em first-order business case} which
does not take transitions into account and a {\em second-order business case} which includes the cost of a transition leading into the sourcement for which the first order business case is specified.

In particular the sourcing history can be of importance only to second-order business cases. The second-order
business case must also contain a risk analysis for the sourcing transition plans.

\subsection{Transformations and transitions}
A transition plan is comparable to a computer program. Its putting into effect produces a progression
of steps which
realizes the intended transformation. More generally a transition description may be compared to a control code \cite{BergstraMiddelburg2009}. A transition effects a transformation. Transformation is a more abstract concept than transition with a particular focus on initial and final states only. A transformation may be understood as the objective of a transition  process which is available before that process is designed in detail; it may also
be viewed as a rationale and in many cases as a summary of the transition plan meant for reporting about the transition
plan after it has been put into effect, with a focus on its outcomes while forgetting the events that took place in between.

In a recursive fashion a classification of dimensions can be proposed which arises
when contemplating transformations and transitions.

\begin{description}
\item{\em Transformation.} Specification of  another (potential, future) state compliant with novel objectives,
given a specification of the current sourcement.
\item{\em Transformational activities.} Transformational activities may involve shipping equipment, hiring and firing of personnel, training of new personnel, knowledge transfer, business process redesign.
\item{\em  Role models for transformational processes.} Roles can be distinguished like: provider, customer, consultant, independent academic professional.
\item{\em Formalized transformational processes.} Transformations will often result from an implementation of a prespecified process intended to effect the required transformation. Transformational processes are also called transitions.%
\footnote{For instance various forms of ETP.}
\end{description}

\section{Attributes at various dimensions}\label{attributes-of-configurations}
A vocabulary of attributes specifies which aspects need to be taken into account at different dimensions. These attributes are best viewed as being attributes for configurations.

\subsection{Attributes of a sourcement}\label{attr-sourcing-configuration}
A sourcement contains information about the facts on the ground. It indicates a family of sources and attributes for these sources. In a (current, past, future, expected, alternative, optimal, suboptimal, adequate etc.) sourcement the following attributes can be distinguished:
\begin{description}
\item{\em Principal unit.} The principal unit of a sourcement
exists besides zero or more secondary or providing units.

\item{\em Theme(s).} Themes were mentioned already in Section \ref{sourceToUnit}. They must be listed once more here in the capacity of an attribute of a sourcement. The theme, or in some cases a coherent group of themes,\footnote{Example: catering, vending machines coffee corners, constitute a coherent group of three themes which may belong to a single principal unit and which may be covered by a single sourcement.} indicates coherent packages of objectives for the principal unit.

\item{\em Sources.} Themes are realized by means of sources which are placed inside units. A sourcement (best accessed via an architecture) displays a mapping from themes to sets of sources (used for achieving the thematic objectives) as well as an assignment of units to sources (indicating the home unit of the source). Thus each source has a unique home unit and may serve for zero or more themes.

\item{\em Thematic operations.} These are carried out by and by means of the sources mentioned in a sourcement.

\item{\em Providing unit (or units).} The providing units or suppliers provide sources contributing to achieving the goals involved in the theme.

\item{\em Location.} Locations for operational work are a specific kind of sources.

\item{\em Facilities and equipment. }These are usually placed at specified locations, where operating staff (employed by one or more units) makes use of them to fulfill the requirements of one or more themes.

\item{\em Operational staff.} These staff members work for the providing unit. The staff may have additional rights of job preservation when follow-up sourcing takes place.
\item{\em Managing staff.} The providing unit managing staff of the operations of which the sourcement consists.
\item{\em Contract management staff.} Contract management resides in the principal unit.
\item{\em Directing staff.}~\footnote{In Dutch directing staff may be translated as `regieorganisatie'.} Directors are employed by both principal unit and providing units as an intermediate team that may support operations in cases where unilateral management is not sufficiently effective.
\item{\em Intellectual property.} Intellectual property is often connected with custom made data, knowledge or software which is essential for the thematic operations.
\item{\em Data, knowledge and software.} These attributes are relevant in as far needed or used when performing the thematic operations. Ownership may be with principal unit, providing unit or otherwise.
\item{\em Stability flag.} The stability flag has two values, stable and unstable. A sourcement has become unstable once it has been decided that it needs to be changed within a specified time frame.  A sourcement is stable from its inception until it becomes unstable. The latter may be a consequence of the progress of time in view of an underlying  contract or it may be caused by a decision made unilaterally by
the management of either the principal unit or of the providing unit.
\item{\em Sourcing history.} The sourcing history of a family of sources in a sourcement explains how it came about, for instance by means of an outsourcing followed by a number of successive follow-up sourcing transformations.
\item{\em Outsourcing history.} Of particular importance is the special case where a sourcing history consists of an initial outsourcing followed by subsequent follow-up outsourcing steps. An outsourcing history may be needed to arrive at an adequate judgement concerning the status of various jobs outside the prinicpal agent.
\end{description}
Principal unit, providing unit(s) and theme are primary attributes. The primary attributes uniquely determine a sourcement within a more complex sourcing equilibrium.

\subsection{Attributes of a business configuration}
Attributes of a business configuration for each unit involved may include the following elements.
\begin{description}
\item{\em Operational options.} Which options are available to parties in order to support their business? For instance, can they perform marketing actions within a unit? And, if so, what form of marketing is permitted? What forms of activity are explicitly not allowed for a provider?
\item{\em Business category.} What kind of business is taking place in economic terms?
\item{\em Profit centers.}  Where are profits made, which streams of products and sales contribute to the results of various parties?
\item{\em Bleeders.} Where are sustained losses expected?
\item{\em Market acquisition motives.} Cases where a provider accepts an uncharacteristic absence of profit or even potential profit. Such cases must be made known and those propelled by such motives must demonstrate that their model is economically sustainable and that delivery is not at risk because the original motive is becoming forgotten.
\end{description}

\subsection{Attributes of a contract configuration}
Many forms of contract can be distinguished for each unit involved:
\begin{description}
\item{\em General law.} Indicates
 relevant aspects that follow from national and international legislation. Such matters should be clarified in individual cases.
\item{\em Rules of trade.} Some relevant constraints may follow from operational rules to which providers have committed autonomously by being member of various organizations.
\item{\em Sustainability charters.} Providers may commit to the delivery of goods and services for which the production complies with one or more international standards concerning ecology, sustainability, fair trade, and, quality of labor conditions.
\item{\em Promises} made by various units may be available in documented form. Promises explain what parties expect of each other
when things are working well.
\item{\em Agreements}  are mutually dependent pairs of promises, covenants are more complex forms of agreements taking
temporal interchange of actions into account.
\item{\em Contracts.} Enforceable agreements formally spelled out in detail optionally including penalties for non-compliance and bonusses for rewarding very good performance.
\end{description}

\subsection{Attributes of a transformation configuration}
\begin{description}
\item{\em Precondition/postcondition.} Pre- and postcondition pairs are informative regarding transformations,
and serve as the input for transition design.
\item{\em Timing and concurrency.} Transformations may be applied in a sequential way for each source, though in parallel for a number of sources.
\item{\em Auxiliary operators.} External forces may be needed to help with enacting a change, ranging from movers to management consultants.
\item{\em Risk analysis.} Each transformation involves risk. Complex risks need to be analyzed.
\end{description}

\section{Outsourcing in detail}\label{Oid}
The terminology and concepts which we have discussed thus far concern states and state
configuration descriptions as well as transformations. Nothing has been said, in more detail than
it was covered in \cite{BV2010} and \cite{BDV2011a,BDV2011b} about the well-known range of
specific types of sourcement transformations which frequently occur in business practice.

Our story line has been motivated from the start by the intention to provide a theory of
outsourcing, and after the preparations made thus far we are in a better
position than before to explain what the different fundamental sourcement transformations,
with outsourcing as the paradigmatic example transformations amount to.

Giving a waterproof definition of outsourcing is hardly doable. The power of the term lies in its flexibility and
in the number of aspects that it aggregates. Nevertheless, by listing sufficiently many of these aspects one comes closer to a useful definition of outsourcing. In doing so a sharper picture is obtained than the one that was presented in \cite{BV2010}.

\subsection{Aspect decomposition of outsourcing}
For a sourcement transformation $F$, which is either planned for the future, or ongoing, or supposed to have taken place in the past, to qualify as an outsourcing the following aspects must be present as essential characteristics, aspects labeled optional may be present:
\begin{description}
\item{\em Acting unit.} Some unit, say $U$, is the actor (agent, principal unit, also called
outsourcing unit, or simply outsourcer) of the outsourcing. This implies that $U$  is:
	\begin{itemize}
	\item Taking the initiative and taking the lead.
	\item Owning (or in charge of) sources or a sourcements which are being outsourced.
	\item Engaging in a transformation towards:
	\begin{itemize}
	\item Making use of a service from some provider $V$, that is becoming a service consumer for a new
	type of service.

	\item Becoming non-selfsourcing for some sources.

	\item Becoming partially or fully non-selfsourcing for some source types.
	Fully non-selfsourcing is 	supposed to apply when all that remains in $U$
	from a presourcement $X$ is a demand management function $X_{dm}$ (which
	may have been newly created at the time of outsourcing and as a side-effect
	of the transition process).

	\item Preparing for being co-value creating with $V$.

	\item Aware of a business case for $V$, in providing the service.

	\item Depending on contracts between $U$ and $V$ regulating $V$'s service provision.

	\item Aware of a termination protocol for the service provision by $V$.
	\end{itemize}
\end{itemize}
\item{\em Single insourcer versus multiple provider/insourcer outsourcing.}
As a partner in the outsourcing transformation $U$ may deal with one or more units $V$, $V_1$,..., $V_n$ often referred to as service providers (or service vendors, or insourcers).
Below the single insourcer case is spelled out in more detail, the multiple insourcer case merely being a natural generalization of it. We will speak of multiple provider/insourcer outsourcing because not all units
$V$, $V_1$,..., $V_n$ must necessarily be insourcing, and some may be exclusively delivering services in cooperation with other units that are insourcing as well.

Although multiple provider/insourcer insourcing seems to be a simple generalization of single
insourcer outsourcing there is a critical difference: it may be necessary to ask different providers/insourcers to
actively engage in non-defensive cooperation and this cooperation may be specified by the outsourcer in quite informal terms, simply because it is impossible to spell out in detail to which problems are to be dealt with
by cooperation. This issue is pertinent in particular if the outsourcer is depending on innovation of the business
process at hand and if that innovation must be turned into a collective responsibility of the entire group of providers/insourcers.%
\footnote{It is a possibility to place the primary responsibity concerning the coordination of the cooperation between different providers/insourcers in the hands of one of these.}

\item{\em Extension of outsourcing.} The question ``what is being outsourced''  must have a clear answer.
Different options are possible of which we will list some important types:%
\footnote{Not all uses of the term ``outsourcing'' are convincing in our view. An interesting
example is found in \cite{Valdman2010} where the philosophical question is posed to what
extent it can be a good thing if a person outsources his/her personal decision making
entirely to another individual, serving as a PC (personal consultant). The paper is quite intriguing but
two remarks can be made: the author makes no attempt to trace the term outsourcing back to previous
literature or to be clear about its meaning. This is not untypical for the way philosophical literature
makes use of imported data. And besides that there is no source being transferred and the use of the
term in this case is not convincing, even if the case is only hypothetical. There is also no indication that
the decision making function will be somehow similar after having been transferred to the PC.
}%
	\begin{itemize}
	\item In the simplest case a source $S$ owned by $X$ is transferred to $V$ and the basic
	sourcement that contains $S$ is modified by updating the name of its owner from $U$ to $V$.
	In this case $S$ is outsourced. It will also be said that the basic sourcement
	containing $S$ is outsourced.
	\item According to \cite{AA2010} a job can be outsourced. This is true if the job is placed in
	another unit together with the person who occupies it.
	\item A job description counts among the facts on the ground. If only the job description
	is transferred to another unit we have a less clear case and the contribution of that move to the
	sourcement  transformation at hand
	being an outsourcing is limited. Below we will capture this kind of distinction by way of the so-called
	degree of outsourcingness.
	\item A collection of sources may be outsourced in one transformation, though perhaps becoming
	owned by different insourcers. Then the sourcement consisting of these sources is outsourced.
	\item If a sourcement $X$ is outsourced then this implies that
	
	\begin{itemize}
	\item some description of $X$ is found, and
	
	\item this description is turned into a pattern, and subsequently
	
	\item that  the pattern is instantiated with new elements.
	\end{itemize}
	\item If one holds the view that sourcement $X$ provides the service of realizing theme $T$
	within unit $U$ it is reasonable to hold that a service is being outsourced.
	\item In some cases a theme $T$ is outsourced. That applies if (i) most or all sourcements $X$
	(entirely or in part residing in $U$) connected to $U$'s 	realization of $T$ are outsourced, and (ii)
	$T$ has become a theme of another unit, say $V$, while still operations concerning $T$ are
	carried out as a service for $U$.
	\end{itemize}
\item{\em Unit commitment (optional).} The outsourcing unit $U$ may be in a situation that it commits itself
in principle to use the complementary insourcer $V$ as a partner as long as such a service providing
partner is needed. This state of affairs is plausible for instance if $V$ is a shared service center for a larger
unit $L$ of which $U$ is a subunit. Then $L$ may oblige $U$ to perform follow-up outsourcing after
contract expiration to $V$ again. If $U$ and $V$ are completely independent there cannot exits (or come into existence) any unit commitment from the side of $U$ in the direction of $V$.

\item{\em Outsourcer compensation (optional).} Outsourcer compensation takes place if $V$ compensates $U$ for
the sourcement that it insources. That action has a value for $V$ because it implements services which $V$
can sell to $U$ as well as to other client units.

A high outsourcer compensation by the insourcer is plausible if the transferred sourcement can either provide
a sustainable profit within $V$ for the duration of $V$'s service contract with $U$, or it can produce a sustainable profit by realizing services for other units than $U$. Outsourcer compensation can exist in different forms. The following transaction types  for outsourcer compensation are plausible.
\begin{description}
\item{\em Single compensating transaction.} $U$ sells the sourcement $X$ (or a part of it) to $V$ and $U$ will not be held
accountable by $V$ if either (i)  the sold part of $X$ creates no value for $V$, or (ii) a forthcoming
 follow-up outsourcing of $X$ by $U$ plays the postsourcement
$X$ or components of it out of $V$'s hands (against the interests of $V$) because
of  source commitments by $U$ towards $X$ or components of $X$.
\item{\em Temporally divided compensating transaction.} $U$ is compensated by $V$ in several phases, where part of the compensation may be cancelled if $X$ gets out of $V$'s hands against $V$'s intentions.
\item{\em Compensating transaction for contract duration.} The outsourcer compensation may be calculated under the assumption that after contract termination  the outsourcing phase from $U$ to $V$ ends in a backsourcing of $X$ from $V$ to $U$.%
\footnote{That case subsumes the case where $X$ escapes from $V$ by way of a combination of backsourcing and subsequent outsourcing, i.e. follow-up outsourcing.}
\end{description}

A complicating factor of outsourcer compensation is that it turns the (singe vendor) outsourcing deal
into a two way transaction. Capturing a two way transaction under the rules of the EU procurement process (ETP) can be a challenge in particular if the agreed outsourcer compensation takes place in stages and depends on how the business proceeds from the perspective of $U$ as well as from the perspective of $V$.

\item{\em Timing.} Outsourcing  results in a business relationship between the outsourcing unit $U$ and one or more insourcing units. This relation may have permanent as well as temporary aspects. The service that
the insourcers (or service providers, or vendors) provide has a temporary aspect and it is planned for a
substantial but bounded period of time, say 5 years. It will be agreed that if either party intends not to
extend the cooperation with the same partners this is announced at least some fixed time interval, say
one year, before the agreed period ends.

\item{\em Underlying life-cycle dependency.} The presourcement which is the target of outsourcing originally plays a role
in supporting some business process $P_u$, which relates to the theme that provides the rationale for the
presourcement's existence within the outsourcing unit's sourcement portfolio. $P_u$ may range from being fairly stable
and indefinite in time (such as various forms of catering,
helpdesk activities, maintenance, cleaning, and administrative support) to being temporary or strongly life-cycle
dependent (such as constructing or demolishing a building or testing new software).  Of course by combining
similar  life-cycle  dependent activities in different phases of their cycle within a single
activity can be obtained that features no life-cycles on average.

If the presourcement supports a process with a marked life-cycle, that is  a life-cycle featuring a succession of rather distinct stages (with classical software engineering as a typical example) then different circumstances need to be distinguished: will sources soon become redundant if a next phase of $P_u$ is reached, or the other way around. In the first case outsourcing
may be motivated by the insourcer's capability to provide additional workforce and in the second case it may
be motivated by an insourcer's capability to accommodate a workforce that increasingly needs to be provided
with new tasks.

\item{\em Service origination.} The outsourcing step introduces a new service. Although this service may
be similar to many services that have been used or designed before its novelty can be derived from the fact that human decision making within $U$ must have been involved when designing the service and
often for providing sources for its realization.
\item{\em Service characteristics.} The service that $V$ offers after insourcing from $U$ can have different
characteristics.%
\footnote{The service characteristics  mentioned here seem to be absent from service science. That makes service science inadequate as an exclusive basis for outsourcing theory.}
These are important in view of the termination protocol of the service.
	\begin{description}
	\item{\em fully source non-committing.} This pertains if at service termination (i.e. the termination of the 	 service contract, not merely of some of its sessions) unit $U$ is in no need or expectation to take 	sources into account which $V$ has made use of for providing the service. In other words $U$ is
	uncommitted to the sources used by $V$ for providing the service realizing theme $T$ of $U$.%
	\footnote{This formulation may be adapted to different wordings of the extension of the
	outsourcing as mentioned before.}
	This case splits in two subcases:
	\begin{description}
	\item{\em Intentionally fully source non-committing.} Both $U$ and $V$ have agreed that no
	source commitments from $U$ will exist or arise after the outsourcing has taken place, and during the
	agreed period of non-selfsourcing for $U$ and service delivery for $V$ thereafter.
	\item{\em unintentionally fully source non-committing.} No agreement has been made but no
	source commitments of $U$ arise either. This is often the case in IT sourcing where any physical
	link between IT staff and $U$'s facilities is likely to be purely accidental and of a far shorter
	duration than 	the agreed period following the outsourcing transformation.
	\end{description}
	\item{\em partially source committing.} (Equivalent to the service being partially
	source non-committing.) This means that some sources%
	\footnote{In practice only sources exclusively consisting of human resources can be the subject
	of a source commitment.}
	 made use of by $V$ are so much connected
	to the processes that run in order to implement theme $T$ of $U$ that at contract termination $U$ is
	not at liberty to forget about the existence of these sources and instead must to some extent
	guarantee that these very sources will be involved in any follow-up outsourcing of $X$.%
	\footnote{Several remarks can be made about source commitment, as a feature
	acquired by $U$ and $V$ together as a consequence of outsourcing from $U$ to $V$.
	
	(i) It may be so that $U$
	requires $V$ to see to it that if $V$ introduces new sources needed
	for its provision of the service, $U$ will not be or become committed to those new sources.
	
	(ii) Source commitment implies constraints for $X$ after contract termination. It does not imply the right
	of $X$ to take command of the sources to which it is committed in case it intends to reposition these
	with another providing unit for a successor contract.
	
	(iii) Source commitment usually results from an insourcing event by $V$ that takes place as a
	constituent of the outsourcing process performed by $U$.
	By insourcing these sources $V$ reduces the cost for $U$ of
	moving towards a state in which its theme $T$ is realized by an external service provider.
	
	(iv) It is unclear to us to what extent the identification of source commitments depends on the history
	of a sourcement. Is it possible to read off all existing source commitments from a current sourcement without
	paying attention to its history? (As noted the sourcement history can explain the existence of the source
	commitments but it may not be needed to define them.)
	}
	\item{\em fully source committing.} This holds in the extreme case that all sources employed
	by $V$
	for the provision of the (newly created) service (with the exclusion of a contract management
	function which may have been created by $V$ for delivering the service, possibly in
	cooperation with $U$'s 	demand management function) are committing for $U$.
	\end{description}
\item{\em Pre- and postconditions.} Outsourcing is a transformation, for that reason its description
involves a specification of facts, business configurations, and contract configurations before and after the
transformation has taken place. The status description before transformation (precondition)
can be explained in terms of
the jargon proposed in previous section. The status description after outsourcing (postcondition)
may be in part
unknown, in particular if vendor selection is part of the outsourcing process.%
\footnote{It is a particular fact of ETP (see \cite{ETP}) based sourcing processes that
when deciding to perform an
outsourcing that decision is made unilaterally by the outsourcer and only a description of one or more so-called lots is available. A definition of a lot is provided in \ref{lotdef} below.
When deciding
against the continuation of a contract the party that has issued that contract through an
outsourcing process may become obliged (in case backsourcing is no option) to perform follow-up outsourcing via ETP while not yet knowing to what extent a multi-vendor follow-up outsourcing will be needed or preferred. The impact of a non-continuation decision is amazing, and the rules implicitly
proposed by the advice for
taking the decision and control factors of \cite{Delen2005} into account can hardly be followed.}
\end{description}

\subsection{Degree of outsourcingness}\label{DOO}
A sourcing transformation between outsourcer $U$ and insourcer $V$ always involves a two-way transaction.
Obviously there are two extreme cases, (i) if only a service is contracted by $U$ from $V$ and no source is transferred  in the other direction, (ii) if only a source is transferred but no service is contracted. In the first case
there we speak of service acquisition (itself to be achieved by implementing an adequate procurement process)
instead of outsourcing and in the second case  we will use the phrase source externalization.

Formally one may split both transactions and view an outsourcing as a combination of a
service acquisition and a source externalization. Both aspects must be somehow of an equal size if a sourcing transformation is to be labeled an outsourcing without hesitation, that is in an ``ideal case''.  Now the ideal case is
less likely to occur and indeed there may be intermediate cases. We will now conceptualize the deviation from the ideal case in terms of a so-called degree of outsourcingness which is assumed to equal 1 in the ideal case. One may
attribute to combinations of a service purchase and a source externalization a degree of outsourcingness
(D$_{outs}$), a
rational number between 0 and 1 which estimates how perfect the transformation matches the concept of an outsourcing. Having available the degree of outsourcing one can speak of intermediate cases. Here are some rules of thumb concerning D$_{outs}$:
\begin{enumerate}
\item If there is no service bought or no source transferred D$_{outs}$=0. This covers both extreme cases mentioned above.

\item If there exists a market for the service and another market for  the source externalization and performing the transactions independently is likely to be at least as economic as performing them simultaneously D$_{outs}$=0.

This may apply if the contracted service is likely to be produced by other sources of $V$ than the  sources it has insourced from $U$ (and that produced the same service within $U$ before their outsourcing).

\item These two rules suggest that D$_{outs}$=0 if the outsourcing can equivalently (or even better) in
financial or economic terms be carried out as a service purchase followed by a source externalization. (In the other order there is likely to be an unacceptable gap in the delivery  of the service.)

\item If the service will be provided initially (say the first 1 or 2 out of 5 contracted years) by exactly the sources that
have been transferred, though somewhat reduced because of $V$'s known (and expected) ability to manage the service production more efficiently and with these (reduced) sources not providing services to other clients of $V$,
D$_{outs}$=0.7 at least.

\item The step from 0.7 to  1.0 can be made if, all of the following conditions are satisfied (is only some are met
D$_{outs}$ will end up between 0.7 and 1.0):
\begin{itemize}
\item The process $P_u$ that underlies the presourcement that $U$ intends to outsource is stable in the sense that
life-cycle effects in its activities average out.
\item The transaction is more economic for $U$ than it would be to, (i) purchase the service and
(ii) dismantle the sourcement (formally the same as externalization to an inactive unit,
a step which is always expensive for $U$).
\item The plausibility that the transferred sources will have been successfully incorporated in $V$ (so that
they may work for other clients if $U$ performs non-trivial follow-up sourcing) after
contract termination is significant.
\item The transfer of sources to $V$ is made possible (in terms of business risk)
for $V$ because of the certainty that the contracted service can be delivered to $U$ for the contracted period.
Otherwise the risk would be too high and the probability of
successful incorporation drops too much as a consequence of the difficult initial phase.
\item In the hands of $V$ the transferred sources (assuming the service contract) are an asset that can be valued
positively (or at least are an asset that can be developed into an asset with positive valuation during the contract period). That positive value is shared between outsourcer and insourcer so that the outsourcer receives a
cheaper service than on a market where its has no corresponding source transfer on offer.
\end{itemize}

\item If the service bought has a  larger volume than what the transferred sources will produce within $V$,
D$_{outs}$ is below 1 because there is a volume of service contracting that is not matched by a complementary
source externalization. The bigger this gap the smaller D$_{outs}$.

\item If the service bought is  smaller in volume than what the transferred sources used to produce before their
outsourcing the sourcement externalization aspect has an overweight and D$_{outs}$ will be below 1.

\item In the case of multi-party outsourcing D$_{outs}$ can be defined by considering the union of the insourcing units
as a singe unit.

\end{enumerate}

If D$_{outs}$ is low then one may consider a decomposition of the transformation as a service purchase (a matter of sourcing) and a source externalization. A considerable complication is that if one splits an outsourcing in a simultaneous service contracting and a source externalization and one attempts to determine prices for both transformations independently, both prices may be positive, zero, or negative, thus leading to nine possible combinations.

An outsourcing can be imagined in advance by specifying a future sourcement and leaving open some of its units. In
that case the number of insourcing units is known and so is how these new units will produce the required services
by making use of the insourced sourcements.

Alternatively an outsourcing can be specified by providing less information and only indicating which
sources should be transferred to other units while leaving open how many insourcing units will be made use of
and leaving room for different kinds of relations between the outsourcing unit and the different insourcing and (thereafter) service providing units. In the latter case the notion of a degree of outsourcingness can be applied in
hindsight only, and a high degree is only attained if the transformation decomposes in a family of sub-outsourcings.
The latter is by no means guaranteed in the case of multi-party outsourcing, however.

In \cite{WibbelsmanMaiero1994} sourcement configurations called co-sourcing are discussed. The transition to co-sourcing is similar to outsourcing but with a lower degree of outsourcingness.

\subsection{Classifying an outsourcing according to resulting service characteristics}
One may distinguish several classes of outsourcing that one finds when focusing on the service characteristics of the
service created in the process of outsourcing. Here it is assumed that the characteristics are phrased
in terms of intentions and that these may differ from facts.

Each of these classifications is about the underlying transition, that is it provides dynamic information. Thus source commitment engaging is about a transition into a sourcement that contains new source commitments.
\begin{description}
\item{\em Source commitment engaging.} This classification
holds in those cases where outsourcing unit $U$ becomes committed to sources to which is was not committed before.%
\footnote{An example may be helpful. If $U$ outsources to $V$
a restaurant (rather the operation of its restaurant which remains  located on $U$'s premises
and has $U$'s clients and employees as customers)
making use of part-time workforce only (the employees working most of their time for different
food and drink delivery services outside $U$), then $V$ may run the business so successfully that,
as time progresses, a number of staff members become full-time employed by $V$ for their work in connection with $U$'s
former restaurant. Now $U$ has become committed to the group of these staff members, a commitment
that is connected to the individuals in group as a source and which may outlive the original
outsourcing and service provision contract.
$U$ may be unhappy about this course of events and take precautionary action
by contractually obliging $V$ to see to it that no such workforce emerges.

Alternatively $U$ may welcome this course of events and consider the commitment as
an opportunity. The latter case may for instance apply if $U$ is more appreciative of $V$ as a
recruiter than as a service provider.}

\item{\em Source commitment preserving.} Before outsourcing initially, $U$ was committed by definition
to its own sources. If the commitments to any of its sources have not degraded as a consequence of the outsourcing transformation
the outsourcing is commitment preserving.%
\footnote{$U$ can be committed to sources not owned by $U$. This can be understood as the result of sourcing decisions. If $V$ offers a service to $U$, that may force $U$ to accept source commitments to sources owned by $V$ before the service delivery was started and before the service contract was signed. This is remarkable and couterintuitive.
Such source commitments may also outlast the duration of the service contract, which is of course the interesting case. This is remarkable: even if all sources used by provider $V$, which
has been selected by $U$ as a pure service provider when pursuing its sourcing strategy,  are owned by $V$, unit
$U$ may
find itself in the situation that it has acquired a source commitment to sources owned by $V$ in such a way
that renewal of the service contract (potentially with another service provider) becomes very much like follow-up outsourcing. Ignoring sources when dealing with services is impossible in general. This observation
seemingly refutes a dogma of service-dominant logic.}
\item{\em Partially source commitment preserving/partially source commitment discharging.} If commitment to some sources has disappeared the outsourcing is partially source commitment
preserving/partially source commitment discharging.
\item{\em Fully source commitment discharging.} If no new (compared with the commitments between $U$ and $V$ before the outsourcing under consideration) source commitments to sources owned by $V$
result as a function of the outsourcing transformation or of its corresponding service provision the outsourcing is fully source commitment discharging.%
\footnote{This seems to be the default assumption in service science.}
An outsourcing may be fully source commitment discharging (intentionally) and the same time it may be
unintentionally (or factually, or accidentally) not fully source non-committing (that is unintentionally partially source committing).

\end{description}

If an outsourcing is fully
source commitment discharging the result is a service usage relation between $U$
and $V$ which in its subsequent development may be dealt with in terms of services only. Outsourcing is
more complex than service acquisition because it can take a variety of source commitments into account. In different words: the language of services
capitalizes on a behavioral abstraction from sourcements to services which is too abstract in many cases
by ignoring sources and source commitments. Especially when a service is terminated this
abstraction has to be undone and outsourcing theory supports just that.

\section{Other fundamental sourcement transformations in detail}\label{Ofstid}
In this section we will consider source externalization, source internalization,  insourcing, backsourcing,
follow-up outsourcing and chained follow-up outsourcing in more detail.

\subsection{Source externalization and source internalization}
The limiting case that sourcement $X$ is outsourced by $U$ to $V$, without $U$ asking $V$ for the
provision any services related to $X$, will be called source externalization. Complementary to source externalization is source internalization, which is $U$'s source externalization seen from the perspective of
$V$.

\subsection{Insourcing}
If one unit is outsourcing one or more other units are insourcing correspondingly. These units are often
referred to as the vendors or as service providers for an outsourcing.%
\footnote{We notice that \cite{BLA2009} defines insourcing as  (the practice of) continued self-sourcing
after due contemplation of the outsourcing option. In contrast to this definition we will
denote with insourcing the symmetric
counterpart of outsourcing, that is the role of the vendor or service provider involved.%
}
Insourcing may be done by initiative of the insourcer if it has some formal power. For instance a new SSC (shared service center) may have been given the power to insource sources, processes and themes from other units. In technical terms
insourcing is quite similar to outsourcing. Indeed most properties of insourcing stem from
the fact that an insourcing of $V$ from $U$ presupposes an outsourcing from $U$ to $V$
to take place simultaneously.
There are some important differences between insourcing and outsourcing worth mentioning.
\begin{itemize}
\item Multivendor insoucing seems not to occur.%
\footnote{One may imagine a simultaneous insourcing from different outsourcing parties. But that
represents multi-seller insourcing rather than multi-vendor insourcing,
because when insourcing takes place the insourcing party plays the role of the vendor (of
the service which is defined during and as a result of the insourcing process).}
\item Multi buyer insourcing is the same as multi party outsourcing but seen  from the perspective of one of the insourcing
parties.
\item Insourcing is not achieved through ETP with the insourcing party in the tendering role. This is an obvious state of affairs if the insourcing unit receives a payment for its promised services, though it is less obvious if the
insourcer pays a substantial sum for the activities it incorporates from the outsourcer.%
\footnote{By definition ETP primarily constrains the initiating unit. A prospective insourcer $V$ is free to offer a bid to another unit for insourcing an attractive sourcement. A reason for doing so may be that $V$ sees
a high potential for achieving economy of scale after the insourcing has been effected.}

\item Insourcing can be carried out by $V$ in theory without affecting its mission statement. Insourcing will not disable mission defining or mission critical operations that $V$ had in place before it decided to perform the insourcing. Thus insourcing need not be mission aware, at least much less than outsourcing.%
\footnote{It is reasonable though to require that an insourcer understands why the complementary outsourcing is considered mission compatible by the outsourcing unit. So the insourcing unit must analyze the whole transaction
in a mission aware style, taking the outsourcing unit's mission into account.}

It is plausible,
however, that insourcing leads to an extension of $V$'s mission statement, which after several steps may
need a stronger profile which, in due time may render other activities non mission co-defining, with the
effect that those become candidates for outsourcing.
\item The term follow-up insourcing seems not to be useful because the insourcer is supposed not to
be in the lead of the insourcing process. If follow-up insourcing stands for finding a new client for
a service after a current contract (which has resulted from insourcing) has been terminated,
the term is ill at place because the intuition of theme ownership being more constant than the
source ownership does not apply.
\end{itemize}

Source commitments are important for insourcing just as well as for outsourcing.

\subsection{Backsourcing}
Backsourcing is one of the options for terminating the service usage that results from outsourcing.%
\footnote{Empirical literature on backsourcing is now emerging (see \cite{BLA2009}). That paper contains
useful further references for the concept of backsourcing.%
}
The intuition is that the effect of the outsourcing transformation which led to a particular sourcement
 is somehow undone. In particular the sources to which a source
commitment from $U$ existed after the outsourcing are insourced again by $U$.

The entity that may be backsourced may be a sourcement, a source, a group of sources, or a theme. This depends on the corresponding variation in outsourcings.
Backsourcing is a particular form of insourcing, except for the fact that a service agreement
is annihilated instead of being created, and except for the fact that the sourcement obtained after
backsourcing is considered permanent. It is more precise to view backsourcing (by $U$, after having outsourced to $V$) as an instance of source internalization by $U$ (from $V$) which also qualifies as a
source externalization by $V$ (to $U$).

If the outsourcing (say of theme $T$) has been
source commitment discharging it is implausible to speak of backsourcing if, after the  termination of the
service provision relation, $U$ internalizes the
provision of the resulting service. A better way to express what takes place is to say that $U$ opts for
becoming self-sourcing again for its realization of the theme $T$. This is a matter of new source
development rather than of backsourcing.%
\footnote{We suggest backservicing as a name for what takes place in this case.}

\subsection{Follow-up outsourcing}
After the service provision and usage phase resulting from an outsourcing (of sourcement $X$ from $U$ to $V$)
has come to an end, termination takes place and many options may be open for carrying out that particular termination.
Backsourcing is an option, and so is discontinuation of the service by the (original) outsourcer. Besides
these options to engage in a follow-up outsourcing is the major alternative strategy.

The simplest way to imagine a follow-up outsourcing is to perform a hypothetical backsourcing
followed by another outsourcing.%
\footnote{This explanation prevents us from saying that  backsourcing is a particular form of follow-up
outsourcing, as a part of the definition of backsourcing. It may be considered
a possible alternative to follow-up outsourcing.}
This explains how source commitments are inherited.
Another way of looking at it is that $V$ performs an outsourcing to some other unit say $W$ (or which is
nowadays more common  to a number of units $W_i, i\in I$ thus constituting a follow-up
multiple outsourcing), in such a way that:
\begin{itemize}
\item $W$ becomes the new service provider (or $W_i, i\in I$ becomes the new service provider group) for $U$ of the service that it accepted from $V$ after its outsourcing of $X$ to $V$ which now comes to an end.
\item $W$, or a unit inside $W$, insources those sources to which $U$ had a source commitment resulting from the outsourcing of $X$. $W$ takes responsibility too for the sources which have been newly created inside $V$ and to which $U$ expects to acquire a source commitment.
\item Business cases and contracts may be renewed and updated.
\item Finally it is required that $V$ and $W$ are comparable in the sense that both may be seen as competing more or less on equal terms on  the market of units that may offer to insource $X$ from $U$. This is clear if the original
outsourcing $X$ could have been outsourced as well to $W$. It may be the case, however, that $W$ came into existence after the outsourcing of $X$ from $U$ to $V$. In that case hypothetical reasoning may be needed to establish the functional comparability of $V$ and $W$.
\end{itemize}

Follow-up outsourcing is non-trivial if $U$ changes its contract partner. Consequently
backsourcing may be considered a non-trivial  case of follow-up outsourcing, but that view is merely and observation, it does not serve definitional purposes.%
\footnote{In case calls for service provision need to be performed by way of ETP, it is plausible to call a
follow-up outsourcing non-trivial if vendor (service provider plus committed sources insourcer) selection is done purely by means of ETP.}

In some cases follow-up outsourcing may be considered an instance of  service contract maintenance.
If outsourcing of $X$ has been fully source commitment discharging the notion of follow-up outsourcing
becomes identical to merely acquiring a new service provider for a known service. In this case service science and the theory and practice of service acquisition brings all methods and tools that are required.

Indeed it may be concluded that after contract termination the next step following an outsourcing need not
be viewed as a follow-up outsourcing but may in some cases be simply considered a renewal of a service contract which may involve the selection of a new service provider. One may understand contract renewal as one of the options available for contract maintenance.

Leaving aspects of unit identity and mission aside follow-up outsourcing seems to reduce to service procurement
after end of service contract while taking source commitments into account that have been acquired by the
service provider during the course of its service provision.%
\footnote{This fails if the original outsourcing the sources transferred to the insourcer were highly valued
and it was agreed that when a subsequent follow-up outsourcing results in a significant part of those sources
becoming taken away from the insourcer (in favor of being insourced by a new business partner of the follow-up outsourcing unit) the insourcer is compensated for that. This mechanism is not a source commitment, rather it
comprises an insurance issued by the outsourcer on behalf of the insourcer against a risk of
non-trivial follow-up outsourcing by which it loses its service contract as well as the sources able to
comply with such contracts.}
 If this is true, the subject of EU procurement for
follow-up outsourcing is subsumed by EU procurement of services. This holds because the issues concerning the
persistence of unit identity and mission have been dealt with in the original outsourcing and need not
be reconsidered in a follow-up outsourcing, assuming that a follow-up outsourcing is pure, that is, (i)
it involves no transfer of sources from the original outsourcer to a new insourcer, and (ii) it involves the
handover of all sources that were originaly insourced from the outsourcing unit and which have been active only or mainly for producing the service which the outsourcing unit began consuming because of the outsourcing transformation.

A follow-up outsourcing may in some cases consist of
a combination of a pure follow-up outsourcing and a relatively minor initial outsourcing both with the same unit as insourcer. It may also consist of a
partial backsourcing followed by a pure follow-up outsourcing.

\subsection{Chained follow-up outsourcing}
After a follow-up outsourcing there will be a termination of contract and a subsequent follow-up outsourcing may be carried out. In principle a chain of follow-up outsourcing transformations can either go on for ever, or
end in  backsourcing, or, when the service that $U$ receives has become obsolete,  or it ends in a source externalization transformation.

\subsection{Progressive outsourcing}
After an outsourcing of sourcement $X$ has taken place from $U$ to $V$ there constituting a postsourcement
$X^{\prime}$, it may be useful for $U$ to effect the externalization of sourcement $X^{\prime}$ to some further unit $W$ in such a way that $W$ will now deliver the service for $U$ that $V$ began delivering upon the outsourcing of $X$ to $V$. In this case we speak of a progressive outsourcing from $U$ to $W$ if the following conditions are met:
\begin{itemize}
\item $V$ stops offering services like the one embodied in $X^{\prime}$ to any other party. In other words it performs
externalization by type.
\item $W$ is a larger unit servicing many more different clients than $V$ with services similar to the one produced by $X^{\prime}$.
\item For $W$ carrying a sourcement like $X^{\prime}$ is more strongly tied to its mission than it is for $V$.
\end{itemize}
Progressive outsourcing (and its chained version mentioned below) is most plausible for sourcements which implement processes without marked life-cycle effects (for instance: cleaning, help-desk support), and educational tasks.
\subsection{Chained progressive and follow-up outsourcing}
Progressive outsourcing can be performed in a number of steps. This is quite visible with the evolution of a shared service center which results from transformations involving intermediate stages. In fact follow-up outsourcing can be used between a chain of progressive outsourcings (or the other way around which is less plausible).

\section{Transformation requirements capture}\label{Trc}
For outsourcing and follow-up outsourcing requirements must be formulated concerning a new sourcement. This
necessitates a requirements capture process. The outcome of that process can be imagined as a sourcement description pattern. After specifying the nature of such patterns a definition can be derived of the notion of a lot, which
is a key feature in any tendering process.

\subsection{Sourcement description patterns}
When outsourcing or follow-up outsourcing must be performed by unit $U$ some sourcements of $U$'s sourcement portfolio must be modified. It would be wrong to say that a sourcement is outsourced itself, rather the process of outsourcing changes a sourcement by perhaps introducing new sources instead of
existing ones, by introducing new owners for existing sources, by removing providing units which will not be made use of anymore.

A way to look at this matter is as follows: before the transformation takes place $U$'s sourcing
architecture is charted. That leads to a description of the various sourcements used by $U$. Now some of the described sourcements must be modified. As a preparation each description of a sourcement which is
expected to change is made into a {\em sourcement description pattern} by replacing known unit names which may be replaced by variables for unit names and by replacing source descriptors for sources that may as a consequence of the intended  transformation.

Sourcement description patters provide architectural information for prospective providers who intend to submit a bid for proving the sources called for. A sourcement description pattern may be viewed as a request for single or combined external units to match the pattern with their own sources (perhaps after insourcing some of the sources of the outsourcing unit $U$).

\subsection{Lots and bids}\label{lotdef}
Sourcements description patterns can be combined into lots. The idea of a lot is that a single and coherent bid is expected for its provision.  A bid indicates a family of closed source descriptions
(that is a pattern with all of its variable elements bound to unit names and source names),
for the patterns listed in the lot.
In a provider selection phrase different bids on the respective lots are compared and a best fit is chosen.
In the subsequent transition phase organizational changes are effected so that the bid becomes a valid
description of the new reality. The process of transition involves actual instantiation of the variables with
units and sources.

\section{Conclusions}\label{Conc}
In this paper the results of \cite{BV2010} and \cite{BDV2011a,BDV2011b} have been extended in three directions: (i) detailed terminology about sources, (ii) the configuration of steady states pertaining between successive sourcing transformations with special attention paid to historic aspects,  (iii) the terminology needed to allow for non-circular descriptions of sourcements, and (iv) a detailed explanation of the mechanisms of the
so-called fundamental sourcing transformations.

Together these developments create an expressive language for discussing the theory and practice of outsourcing which allows for modularity and for a stratification of concepts. Considering the stratification the most important
aspect of the development of this paper we will refer to its outcome as a stratified outsourcing theory. As a theory its main purpose is to clarify and disambiguate terminology and from that viewpoint to survey the many different
aspects of sourcing transformations in a systematic way. The outsourcing theory outlined above is not aiming
at an explanation of empirical observations. Nevertheless extracting some empirical content from it may become
possible on the long run.


\begin{thebibliography}{99}
\bibitem{AA2010}
M. K.\ Andersen and P. Ankerstjerne.
\newblock HR issues to be considered when outsourcing services.
\newblock {\em Aspector/ISS ISS white paper.} ({\tt www.aspector.dk}) (2010).

\bibitem{BaetenBastenReniers2009}
J.C.M.\ Baeten, T.\ Basten and M.A.\ Reniers.
\newblock Process Algebra: Equational Theories of Communicating Processes.
\newblock {\em Cambridge Tracts in Theoretical Computer Science.} Vol. 50, (2009).

\bibitem{BallantyneVarey2008}
D.\ Ballantyne, and R.J.\ Varey.
\newblock The service-dominant logic and the future of marketing.
\newblock{Journal of the Academy of Marketing Science.} Vol 36 pp. 11-14 (2008).

\bibitem{BLA2009}
H.T.\ Barney, G.C.\ Low, and A.\ Aurum.
\newblock The morning after: what happens when outsourcing relationships end?
\newblock{\em In: G.A. Papadopoulos et. al. eds. Information Systems Development,
Springer LLC} pp. 637-644 (2009).

\bibitem{Bergstra2010a}
J.A.\ Bergstra.
\newblock Formaleuros, formalcoins and virtual monies.
\newblock 2010.
\newblock {\tt arXiv:1008.0616 [cs.CY]}.

\bibitem{BDV2011a}
J.A.\ Bergstra, G.P.A.J.\ Delen, and S.F.M.\ van Vlijmen.
\newblock Introducing Sourcements.
\newblock  {\tt arXiv:1107.4684 [cs.SE]}, (2011).

\bibitem{BDV2011b}
J.A.\ Bergstra, G.P.A.J.\ Delen, and S.F.M.\ van Vlijmen.
\newblock Outsourcing Competence.
\newblock {\tt arXiv:1109.6536 [cs.OH]}, (2011).

\bibitem{BergstraMiddelburg2007}
J.A.\ Bergstra and C.A.\ Middelburg.
\newblock Thread algebra for strategic interleaving.
\newblock {\em Formal Aspects of Computing}, 19 (4):445--474, (2007).

\bibitem{BergstraMiddelburg2009}
J.A.\ Bergstra and C.A.\ Middelburg.
\newblock Machine structure oriented control code logic.
\newblock {\em Acta Informatica}, 5(1):170--192, (2009).
\newblock {(\tt arXiv:0711.0836 [cs.SE])}.

\bibitem{BergstraMiddelburg2011}
J.A.\ Bergstra and C.A.\ Middelburg.
\newblock An application specific logic for interest prohibition theory.
\newblock {(\tt arXiv:1104.0308 [q-fin.GN])}, (2011).


\bibitem{BV2010}
J.A.\ Bergstra and S.F.M.\ van Vlijmen.
\newblock Business mereology, imaginative definitions of outsourcing and insourcing transformations.
\newblock  {\tt arXiv:1012.5739 [cs.SE]}, (2010).

\bibitem{Beulen2000}
E.P.\ Beulen.
\newblock Beheersing van IT-outsourcingrelaties.
\newblock {\em PhD Thesis, University of Tilburg},  (in Dutch, 2010).

\bibitem{Burgess2007}
M. Burgess.
\newblock System administration and the scientific method.
\newblock in: J.A.Bergstra and M. Burgess (editors), {\em Handbook of Network and System Administration}:
\newblock pp. 689-728, 2007.


\bibitem{Delen2005}
G.P.A.J.\ Delen.
\newblock Decision- en controlfactoren voor sourcing van IT.
\newblock {\em PhD Thesis, University of Amsterdam}, Van Haren Publishing Zaltbommel (in Dutch), (2005).

\bibitem{Delen2007}
G.P.A.J.\ Delen.
\newblock Decision and Control Factors for IT-sourcing.
\newblock in: J. A.Bergstra and M. Burgess (editors), {\em Handbook of Network and System administration}:
\newblock pp. 929-946, (2007).

\bibitem{ETP}
European Tender Procedure,
\newblock http://ec.europa.eu/youreurope/business/ \\profiting-from-eu-market/benefiting-from-public-contracts/index\_en.htm.



\bibitem{KernWillcocks1999}
T.\ Kern and L.P.\ Willcocks.
\newblock Exploring information technology outsourcing relationships: theory and practice.
\newblock {\em Erasmus University Rotterdam}, management report no. 61-1999, (1999).

\bibitem{KimblerBouma1995}
K.\ Kimbler, L.G.\ Bouma (Eds.)
\newblock Feature interactions in telecommunications and software systems V.
\newblock {IOS Press}, 1998.

\bibitem{Lacity1993}
M.C.\ Lacity.
\newblock Information systems outsourcing: myths, methaphors and realities.
\newblock {\em Wiley, Chishester}, (1993).

\bibitem{LohVenkatraman1992}
L.\ Loh and N.\ Venkatraman.
\newblock Diffusion of information technology outsourcing, influence sources and the Kodak effect.
\newblock {\em Information Systems research}, 4: 334--358, (1992).

\bibitem{Middelburg2010a}
C.A.\ Middelburg.
\newblock Searching publications on operating systems.
\newblock {\tt arXiv:1003.5525 [cs.OS]}, (2010).

\bibitem{Middelburg2010b}
C.A.\ Middelburg.
\newblock Searching publications on software testing.
\newblock {\tt arXiv:1008.2647 [cs.SE]}, (2010).



\bibitem{PON2006}
Platform Outsourcing Nederland, werkgroep Taxonomie Outsourcing.
\newblock Outsourcing van IT -- Management guide.
\newblock Van Haren Publishing, (2006).


%
%
%

\bibitem{Valdman2010}
M.\ Valdman.
\newblock Outsourcing Self-Government.
\newblock {\em Ethics}, Vol. 120. (4) pp.761-790 (2010).

\bibitem{VargoLusch2004}
S.\ Vargo and R.F.\ Lusch.
\newblock Evolving to a new dominant logic for marketing.
\newblock {\em Journal of Marketing} 68 pp. 1-17 (2004).

\bibitem{Varzi2009}
A. Varzi.
\newblock Mereology.
\newblock {\em Stanford Encyclopedia of Philosophy},
\newblock {\tt http://plato.stanford.edu/entries/mereology}, (2009).

%

\bibitem{WibbelsmanMaiero1994}
D. Wibbelsman and T. Maieiro.
\newblock Co-sourcing.
\newblock {\em Proc. Outsourcing, Co-sourcing and Insourcing Conference}, Berkely (1994).

\end{thebibliography}
\end{document}